\documentclass[11pt,twoside]{article}
\usepackage{asp2006}
\usepackage{epsf}
\usepackage{psfig}
\usepackage{graphics}
\usepackage{lscape}
\begin{document} 
\title{A Splinter Session on the Thorny Problem of Stellar Ages}

\vspace*{-0.5cm} 
\author{Eric E. Mamajek\altaffilmark{1}, 
David Barrado y Navascu\'{e}s\altaffilmark{2}, 
Sofia Randich\altaffilmark{3},
Eric L.~N. Jensen\altaffilmark{4},
Patrick A. Young\altaffilmark{5,6},
Andrea Miglio\altaffilmark{7}, 
Sydney A. Barnes\altaffilmark{8}}

\altaffiltext{1}{Harvard-Smithsonian Center for Astrophysics, Cambridge, MA 02138}
\altaffiltext{2}{Laboratorio de Astrof\'{i}sica Espacial y F\'{i}sica Fundamental, Apdo. 50727, 28080 Madrid, Spain}
\altaffiltext{3}{INAF-Osservatorio Astrofisico di Arcetri, L.go
E. Fermi, 5, 50125 Firenze, Italy} 
\altaffiltext{4}{Swarthmore College, Dept. of Physics \& Astronomy, Swarthmore, PA 19081}
\altaffiltext{5}{Theoretical Division, Los Alamos National Laboratory, Los Alamos, NM 87545}
\altaffiltext{6}{Steward Observatory, University of Arizona, Tucson, AZ 85721}
\altaffiltext{7}{Inst.
d'Astrophys. et de G\'{e}ophys. de l'Universit\'{e} de
Li\'{e}ge, All\'{e}e du 6 Ao\^{u}t, 17 B-4000 Li\'{e}ge 1, Belgium}
\altaffiltext{8}{Lowell Observatory, 1400 W. Mars Hill Rd., Flagstaff,
AZ 86001}

\begin{abstract} Accurate stellar ages remain one of the most poorly
constrained, but most desired, astronomical quantities.  Here we
briefly summarize some recent efforts to improve the stellar age scale
from a subset of talks from the ``Stellar Ages'' splinter session at
the {\it 14th Cambridge Workshop on Cool Stars, Stellar Systems, and
the Sun}. The topics discussed include both the apparent successes and
alarming discrepancies in using Li depletion to age-date clusters,
sources of uncertainty in ages due to input physics in evolutionary
models, and recent results from asteroseismology and gyrochronology.
\end{abstract}

\vspace{-1cm}
\section{Introduction}

``{\it What is the age of [star catalog] [index \#]?}'' is probably
one of the most frustrating astronomical inquiries that one can make,
and also one of the most quixotic of tasks to tackle.  For studies of
the evolution of interesting properties related to stars
(e.g. planeticity, circumstellar disk evolution, etc.), astronomers
would ideally like a robust means of estimating the age of any star in
the sky. There are difficulties with age estimation throughout the
stellar mass spectrum, with some stellar mass and age regimes more
amenable to age-dating than others due to rapid structural evolution.
To make matters worse for the estimation of ages for field stars, most
useful age diagnostics can manifest a wide range of values amongst
stars in supposedly coeval clusters. There are age-dating methods that
are becoming more {\it precise} (e.g. Li depletion boundary,
gyrochronology, asteroseismology), but it remains to be seen how {\it
accurate} they are when some underlying critical physics and
parameters (e.g. the solar metal fraction, convection, etc.)  are
still matters of active debate.  Of course, the ``golden spike'' of
the stellar age scale is the Sun, which has its age constrained by
lead isotopic ages of the oldest portions of meteorites
\citep[the Ca-Al-rich inclusions; 4567.2\,$\pm$\,0.6 Myr old; ][]{Amelin02}.

At the {\it Cool Stars 14} meeting held in Pasadena in 2006, a
splinter session of talks was held devoted to discussing contemporary
issues regarding the ages of cool stars.  Here we summarize a
subsample of those talks (which due to time constraints covered only a
subsample of age-related topics) and point the reader to recent
studies relevant to the stellar age scale. The text in the following
sections reflects the current state of the field in the view of the
various authors, and no attempt has been made to reconcile their
results and opinions.

\vspace*{-0.4cm}
\section{Stellar Ages: How Far Can We Go? (Barrado)}

A survey in the literature will easily show that, for a given stellar
association, there is a large range of age estimates. These
differences arise from different facts. First, we have different sets
of theoretical models which might produce different answers for the
same dataset. In any case, the comparison between observational
properties and those predicted by theory is not an easy task
\citep[see ][]{Stauffer95}. Moreover, stars are individuals, and they
have properties, such as starspots, accretion rates, initial
conditions, etc., which can alter the age estimate. As an example,
colors can be modified by surface inhomogeneities \citep{Stauffer03}
which can affect a cluster's placement in a color-magnitude diagram.
An additional problem is coevality. It is normally assumed that
members of an association are born almost at the same time, but this
is not necessarily true.  On the other hand, among the methods we use
to estimate ages, both {\it primary} (upper Main sequence and
isochrone fitting, eclipsing binaries, spectral features, lithium) and
{\it secondary} (stellar activity, rotation, group membership), there
are many pitfalls since there are uncertainties in the input physics.
Moreover, each of these methods is valid for a given range of stellar
ages and masses, a fact that is not always honored.  All these sources
of uncertainty make the stellar chronology very uncertain although
sometimes we can find formal errors of tenths of Myr, which are far
from our present capabilities.

From my point of view, we should aim for consistency. A given property
(or properties) allow us to sort a set of stellar associations from
the youngest to the oldest, even if we can not derive absolute
ages. Even if other properties produce a different age scale, at least
we should be able to have the same kind of sequence, with Taurus being
younger than the $\lambda$ Ori cluster, which is younger than
IC\,2391, and still much younger than the Pleiades.  Eventually, we
should be able to construct different ages scales which should be
consistent with each other, and should produce absolute and well as
relative ages.  This only can be achieved with a considerable effort
in the theoretical and observational sides of the
problem. Specifically, we should gather very large datasets with the
same setup for a large number of stellar association, in order to
avoid biases.

\begin{figure}[!ht]
\plotone{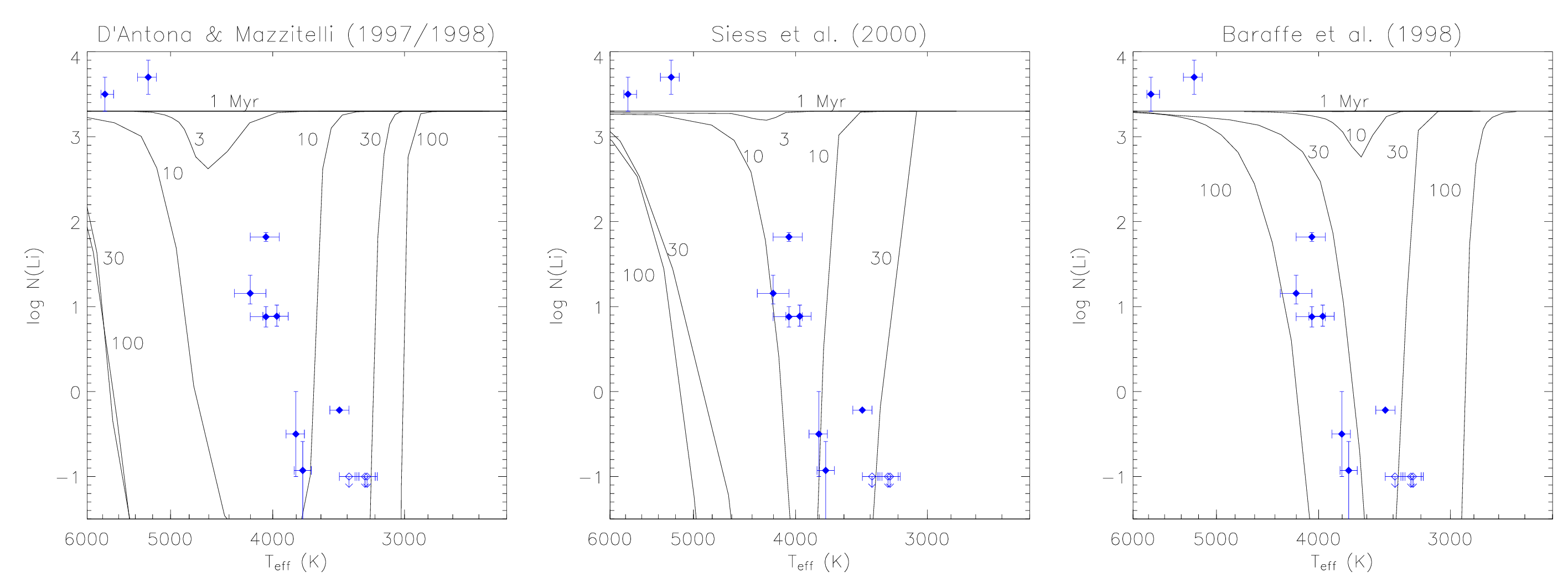}
\caption{$T_{\rm eff}$ vs. Li abundance for stars in the
10--12 Myr-old $\beta$ Pic moving group. For each model, isochrones
for 1, 3, 10, 30, and 100 Myr are plotted.\label{fig:li_depletion1}}
\end{figure}

\vspace*{-0.4cm}
\section{The Li Depletion Boundary in the $\beta$ Pic Group (Jensen)}

Pre--main-sequence (PMS) Li depletion has the potential to be a
valuable age indicator for late-type PMS stars, \citep[see e.g., the
review by ][]{Jeffries00}.  Like any age indicator, however, the
models that tie a particular Li abundance and effective temperature
($T_{\rm eff}$) to an age must be calibrated with empirical data.

We observed twelve late-type members of the $\beta$ Pic moving group
(BPMG) at spectral resolution $R = 40,000$ using the echelle
spectrograph on the 4-meter Blanco telescope at CTIO. Spectral types
were determined from the TiO band indices, and converted to $T_{\rm
eff}$ using the intermediate scale given by \citet{Luhman99} for the M
stars and the main-sequence (MS) scale from \citet{Kenyon95} for earlier
types.  We determined Li abundances by comparing the observed spectra
to a grid of high-resolution synthetic spectra.

Fig. \ref{fig:li_depletion1} shows Li abundance data for the twelve
stars overlaid on three sets of models: \citep{D'Antona97,
D'Antona98,Siess00,Baraffe98}. These stars all have ages of
$\sim$10--20 Myr based on their HR diagram positions, but the
systematic trend observed in their Li depletion is that {\it full
depletion of Li occurs much more rapidly for the M stars than the
models predict}.  In each of the models, the right side of the dip in
the lithium depletion isochrones, which indicates the lithium
depletion boundary (LDB), consistently falls at effective temperatures
too high for our data if we assume the HR diagram ages are correct.
Three stars with spectral types of M4 show fully-depleted Li, with an
upper limit of $\log A(Li) < -1$; none of the models predicts
significant Li depletion in stars of this $T_{\rm eff}$ until ages $>
30$ Myr.

Some authors have questioned whether the BPMG is truly a coeval group
of stars with a common origin.  Even if one accepts an age spread in
the sample, however, the components of individual binary or multiple
systems should be coeval with each other.  Examination of the Li
abundances in individual systems in the sample shows the same
discrepancies as the sample as a whole; the pair BD$-$17 6128 A/B, and
the triple systems HD 155555 A/B/C and GJ 803/799N/799S, are
inconsistent with having a single Li-depletion age for all components.
The direction of the systematic discrepancies reported here are
consistent with previous work showing that the LDB occurs at a cooler
$T_{\rm eff}$ for a given age than the models predict.  Our results
are also consistent with the work of \citet{White05}, who suggest that
the M3 classical T Tauri star could be a young member of Tau-Aur
despite its large Li depletion.

\vspace*{-0.5cm}
\section{Lithium Depletion Ages and Age Spreads (Randich)}

HR diagrams of several young clusters and associations suggest the
presence of age spreads among cluster members. Understanding whether
these spreads are intrinsic or rather due to errors in the
determination of stellar parameters and misplacement of the stars on
the HR diagram, is crucial for gaining insight into the cluster star
formation (SF) histories and, in particular, to put constraints on the
overall duration of the SF process (Hillenbrand, this
volume). Constraints on aspects of SF and early (sub)stellar evolution
can also be gathered through the investigation of PMS clusters with
ages $\sim$10-50\,Myr.  SF in PMS clusters has ceased, implying that
their members have attained their final masses and that these clusters
do not present the complications that are intrinsic to SF
environments. On the other hand, PMS clusters are generally too young
to be dynamically relaxed and thus reveal a cleared view of the final
product of the SF process.  In this context, using the multiplex
instrument FLAMES on VLT/UT2, we have obtained medium/high resolution
spectra of low-mass members of the Orion Nebula Cluster (ONC), the
$\sigma$~Ori cluster, and the PMS cluster IC~4665
\citep{deWit06}. Our main goal was the measurement of their lithium
content, which provides an independent clock to derive the ages and
age spreads.

\noindent {\bf ONC and $\sigma$~Ori --
\citep{Palla05,Palla07,Sacco07}:} We have observed 141 high--probability 
cluster members (membership probability from proper motion
$\geq$90\,\%) down to $\sim$0.1\,M$_{\odot}$ in the ONC and 98 cluster
candidates in $\sigma$~Ori.  Li abundances were derived from veiling
corrected pseudo-equivalent widths (pEWs) of the Li~{\sc i} 670.8\,nm
feature.  Radial velocities were also measured, along with the
strength of the He~667.8\,nm (ONC) and H$\alpha$ ($\sigma$~Ori)
emission lines. All the ONC targets were confirmed as members, while
in $\sigma$\,Ori 59 {\it bona fide} members were identified.  Li
abundances are consistent with the interstellar value for most of the
sample stars in both clusters. However, six stars in the ONC and three
stars in $\sigma$\,Ori show Li depletion factors between $\sim$2-50 (ONC)
and $\geq$100 ($\sigma$~Ori).  Nuclear ages and masses for four
Li-depleted stars in the ONC and two of the three Li-depleted stars in
$\sigma$~Ori are in excellent agreement with isochronal ages and
masses. Inferred ages are in the range $\sim$10-30\,Myr, much older
than the bulk of the cluster populations. Hence, for both regions,
there is independent evidence for a population of old stars mixed with
the much larger assembly of young objects, supporting the suggestion
that SF in the two clusters has not occurred in a short single burst,
but has rather continued for a long time, much in excess of the
dynamical time scale.

\begin{figure}[!ht]
\plotfiddle{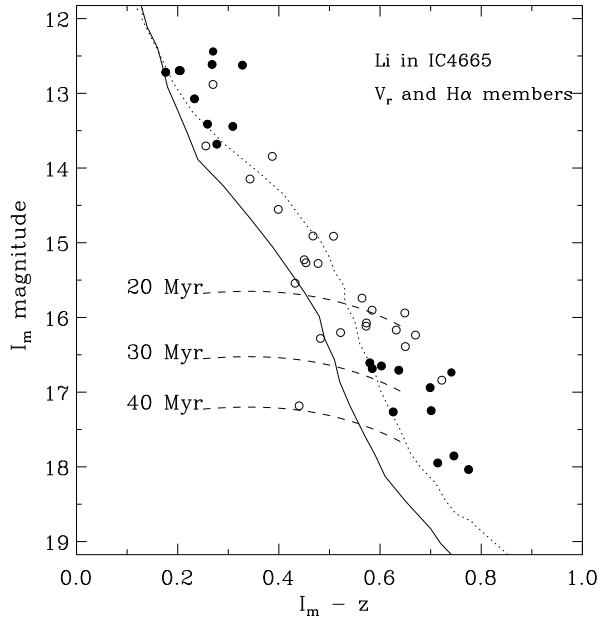}{3in}{0}{50}{50}{-160}{-50}
\caption{I$_{\rm m}$ vs. (I$_{\rm m}-z$) diagram for members of
IC\,4665. Filled and open symbols indicate stars with and without
Li. The 100 Myr (solid line) and 30 Myr (short-dashed line) isochrones
from \citet{Baraffe98} are also shown, along with the predicted LDBs
for 20, 30 and 40 Myr. \label{fig:randich_ldb_ic4665}}
\end{figure}

\noindent {\bf IC~4665 --(Manzi et al. in prep.):} With a
turn-off age of 36~Myr, this cluster represents a good candidate PMS
cluster. We observed 158 candidate members and confirmed membership
for 45 of them based on radial velocity, presence of H$\alpha$ and/or
Li absorption line (depending on mass). The I$_{\rm m}$ vs. I$_{\rm
m}$-z color-magnitude diagram of the confirmed members is shown in
Fig. \ref{fig:randich_ldb_ic4665}, together with isochrones and the
predicted location of the LDB from the evolutionary tracks of
\citet{Baraffe98}. The figure shows a sharp boundary: the
faintest star without a detected Li line has I$_{\rm m}=16.39$, while
the brightest star with Li has I=16.61. The LDB was determined as the
mean of the two values, assuming $d$\,=\,370\,pc, and reddening
A$_{I_{\rm C}}=0.326$. After considering all of the observational
uncertainties, this yields M$_{\rm I_{\rm C}}=8.52 \pm 0.40$.
Accordingly, IC~4665 has an LDB age of $t_{\rm LDB}=27 \pm 5$~Myr.
This value confirms the PMS nature of IC~4665 and agrees with the
turn-off age, the first secure instance for a PMS cluster.

\section{Theoretical Uncertainties in Evolutionary Tracks and Ages (Young)}

The accuracy of stellar ages determined from fitting to evolutionary
tracks or isochrones is limited by the accuracy of the tracks
themselves. Inaccuracies can arise from missing or incorrect physics
in the models as well as physical quantities known with limited
precision, such as metallicity. Some of the most important sources of
error are discussed here individually, but it is important to remember
that there are cross terms between all of these quantities.

The most important problem is the hydrodynamics in stellar
models. Because evolution calculations are treated as a series of
snapshots of static states, dynamics are often ignored. Some elements
such as rotation have been added, but are generally treated
analytically and in isolation. Deficiencies are often corrected by the
inclusion of free parameters calibrated by observations. A calibration
appropriate for one stellar type is generally incorrect for others. As
a result, mixing in stars in very poorly described and not predictive.

Convection is defined by thermodynamic gradients, which is appropriate
only for the onset of convection. Once convection is established, the
dynamically unstable part of the star is larger than the
thermodynamically unstable region, and other processes besides
convection can contribute to mixing. Evaluating the stability of the
stellar stratification against all the available sources of kinetic
energy, including turbulent convection, waves, and bulk rotational
flows gives a self-consistent and predictive description of the
hydrodynamics. Differences from standard models can be large,
especially for massive stars. Larger mixed regions result in higher
luminosities, larger radii, efficient angular momentum transport, and
transport of chemical species to the photosphere. On the MS
model ages for a sample of eclipsing binaries from 1 to 23\
$M_{\odot}$ change by 15--400\%. Radiative transport can be
important for stars with deep convective envelopes. The diffusion time
of energy from a fluid parcel can become comparable or even less than
the crossing time of the convection zone. Convection becomes very
inefficient, leading to larger radii, and ages can differ by 10's of \%.

Even complete and correctly implemented physics can cause problems due
to the inherent lack of precision in our determination of physical
quantities. A comparison of the OPAL and \citet{Timmes99} equations of
state (EOS) gives surprising results. The two formulations differ by
at most 2\% under any of the conditions tested, yet they produce very
different radii on the Hayashi track. Ages determined from each model
can vary by $\sim$60\% or so. The main sequences are not reconcilable
without changing the composition or convective treatment. This is
likely the largest source of uncertainty for sub-solar mass tracks.

Composition, unsurprisingly, plays a significant role. For a 10\%
change in composition, deeply convective stars can have ages different
by $\sim$10\%, while MS ages can change by factors of
several. Given that determinations of the solar composition have
changed by roughly a third in five years, this is a serious problem,
especially when precise metallicities are unavailable for most stars.

The conversion to observables adds another layer of complexity beyond
the scope of this discussion. Model atmospheres, temperature to color
conversions, rotational evolution (which require magnetohydrodynamics
to model), and correct distances are all necessary.

So far we have discussed these sources in isolation. When matching
model tracks, radius is the primary comparison, whether directly or
through $T_{eff}$. Individual effects drive radius in opposite
directions, and so cannot be calibrated out. As a very simplified
example of this interplay, consider a change in the equation of
state. This changes the stratification of the star. A change in the
stratification changes the extent and vigor of
convection. Hydrodynamics enters to change the depth of convection and
introduces new mixing processes that change the compositional
structure and thus stratification of the star. These processes also
transport angular momentum, which feeds back on the stability and
mixing. These change shift opacities as well as solutions for the
EOS. The problem must be solved self-consistently in order to reduce
the uncertainties and produce an accurate result.

In short, hydrodynamics introduces changes of 10's to 100's of \%
in age determinations. Radiation transport adds 10's of \% on the
Hayashi track. The EOS is a surprisingly large contributor at 10's of
\% at 1\,$M_{\odot}$. Composition can be a large effect for relatively
small changes. All of these sources influence each other. These cross
terms can result in much larger synergistic errors, or, more
insidiously, the right answers for the wrong reasons. Fortunately,
multidimensional simulations are allowing us to understand and
generalize the hydrodynamic processes in stars, resulting in a
considerable improvement in the models.

\vspace*{-0.5cm}
\section{Asteroseismological Constraints on Ages (Miglio)}

Though the potential of solar-like oscillations as an effective tool
to determine stellar ages has been known for years
\citep[e.g.][]{Gough95}, it is just recently that, thanks to the
improvement of observational techniques \citetext{see Bedding, this
volume}, the seismic modeling of bright nearby stars has become
reality.

The uniqueness of a seismic inference of age resides in the fact that
oscillation frequencies carry direct information on the structure of
the central regions of the star.  A well-known indicator of stellar
age for MS stars is the so-called small frequency separation
($\delta\nu$) which, as shown by the asymptotic theory of stellar
oscillations \citep{Tassoul80}, represents a probe of the sound speed
in the energy-generating core.  The evolutionary state of a star can
also be inferred by other suitable combinations of frequencies
\citep[see e.g.][]{Roxburgh03} or, in the case of models of evolved
stars, by the detection in the oscillation spectrum of modes of mixed
pressure and gravity character \citep[see e.g.][]{JCD95}.

As an example of one of the first attempts to model solar-like stars
including seismic constraints, we consider the visual binary system
$\alpha$ Cen that, thanks to the combination of precise ``classical''
constraints (masses, radii, chemical composition, luminosity) and the
detection of solar-like oscillations in both components
\citep{Bouchy02,Carrier03,Bedding04,Kjeldsen05}, has recently been the
subject of several theoretical studies \citep[see
e.g.][]{Thevenin02,Thoul03,Eggenberger04,Miglio05}.  As predicted by
\citet{Brown94}, the inclusion of $\delta\nu$ in the modeling of the
system allows one to significantly reduce the estimated uncertainty on
the age of $\alpha$ Cen. If the asteroseismic constraints are not
included in the fit, a value of $8.9 \pm 1.9$ Gyr is found, whereas
the inclusion of seismic constraints reduces the estimated age (and
its uncertainty) to $5.8 \pm 0.2$ Gyr \citep[see][]{Miglio05}.

The accuracy of such age estimates is limited, however, by the poor
frequency resolution of current seismic data and, more importantly, by
uncertainties in the physics included in the stellar models.
Nonetheless, once precise seismic data will be available in
observationally well-constrained targets (e.g. visual binaries),
asteroseismology will also be able to go beyond the determination of
global parameters, and directly test stellar models -- not only adding
precision, but also accuracy to the age estimates.

\vspace*{-0.5cm}
\section{Gyrochronology (Barnes)}  

That the rotation of cool stars slows as a function of age was first
put on a firm observational footing by \citet{Skumanich72}. The wealth
of observational data for cool stars in clusters with ``known'' ages
gathered over the past decade can be used to construct a useful age
estimator from measured stellar rotation periods (``{\it gyrochronology}'').
More details about this can be found in \citet{Barnes03} and Barnes
(submitted to ApJ).

Beyond an age of $\sim$100\,Myr, the rotation rate of any late-F to
early-M star (those with surface convection zones and interior
radiative zones) is describable by $P = f(B-V) \times g(t)$, where $P$
is the rotation period of the star, and $f$ and $g$ are separable
functions of stellar color, $B-V$, and age, $t$, respectively.  Both
$f(B-V)$ and $g(t)$ can be determined purely from the data.  From the
open cluster data, one can show that $f(B-V) = (0.773 \pm 0.011)
\times (B-V_0 - 0.4)^{0.601 \pm 0.024})$.  $g(t)$ can be shown to be
roughly Skumanich \citep{Skumanich72}, and by requiring that the solar
rotation rate be reproduced at solar age, one finds that $g(t) =
t^{0.519 \pm 0.007}$. The preceding considerations allow us to write
the ``gyro age'' of a star as: $log(t_{gyro}) = \frac{1}{n}$ $\{log(P)
- log(a) - b \times log(B-V-0.4) \}$, where $n=0.519 \pm 0.007$, $a =
0.773 \pm 0.011$, and $b=0.601 \pm 0.024$. One can also derive an
error on the gyro age. The fractional error is given by $ \frac{\delta
t}{t} = 2\% \times \sqrt{ 4.5 + \frac{1}{2} (ln\,t)^2 + 2\,P^{0.6} +
(\frac{0.6}{x})^2} $, where $x= B-V-0.4$.  For late-F to early-M stars
of $\sim$1\,Gyr, this works out to be $\sim$15\%.

This method satisfies all five criteria desired for an age indicator: of
finding a variable that changes regularly and sensitively with age, of
being able to calibrate that variable against a very well-defined age,
or set thereof, of being able to identify and measure the functional
form of the variable against its dependent variables (this one is even
separable), of being able to invert the dependence to find the
dependence of age on the other variables, and finally, of being able
to calculate the error as a function of the variables.

\acknowledgements 

The authors thank L. Hillenbrand and D. Soderblom for co-chairing the
session, D. Soderblom and F. Palla for providing closing comments at
the end of the splinter session, and J. Stauffer and the SOC for
supporting the proposed splinter session topic.


\begin{thebibliography}{}
\bibitem[Amelin et al.(2002)]{Amelin02}
Amelin, Y., Krot, A.~N., Hutcheon, I.~D., \& Ulyanov, A.~A.\ 2002, Science, 297, 1678
\bibitem[Baraffe et al.(1998)]{Baraffe98}
Baraffe, I., Chabrier, G., Allard, F., Hauschildt, P.H. 1998, A\&A, 337, 403
\bibitem[Barnes(2003)]{Barnes03} 
Barnes, S.~A.\ 2003, \apj, 586, 464 
\bibitem[Bedding et al.(2004)]{Bedding04}
Bedding T.~R., et al. 2004, ApJ, 614, 380
\bibitem[Bouchy \& Carrier(2002)]{Bouchy02}
Bouchy F. \& Carrier F. 2002, A\&A, 390, 205
\bibitem[Brown et al.(1994)]{Brown94}
Brown T.~M., et al. 1994, ApJ, 427, 1013
\bibitem[Carrier \& Bourban(2003)]{Carrier03}
Carrier F., Bourban G., 2003, A\&A, 406, L23
\bibitem[Christensen-Dalsgaard, Bedding, \& Kjeldsen(1995)]{JCD95}
Christensen-Dalsgaard J., Bedding T.~R., Kjeldsen H., 1995, ApJ, 443, L29
\bibitem[D'Antona \& Mazzitelli(1997)]{D'Antona97}
D'Antona, F., \& Mazzitelli, I.\ 1997, Mem. Soc. Astro. It., 68, 807
\bibitem[D'Antona \& Mazzitelli(1998)]{D'Antona98}
D'Antona, F., \& Mazzitelli, I.\ 1998, ASPC, 134, 442
\bibitem[Eggenberger et al.(2004)]{Eggenberger04}
Eggenberger P., et al. 2004, A\&A, 417, 235
\bibitem[Gough(1995)]{Gough95}
Gough D.~O. 1995, ASPC, 76, 551
\bibitem[Jeffries(2000)]{Jeffries00} 
Jeffries, R.~D.\ 2000, ASPC, 198, 245
\bibitem[Kenyon \& Hartmann(1995)]{Kenyon95}
Kenyon, S.~J., \& Hartmann, L.\ 1995, \apjs, 101, 117
\bibitem[Kjeldsen et al.(2005)]{Kjeldsen05}
Kjeldsen H., et al., 2005, ApJ, 635, 1281
\bibitem[Luhman (1999)]{Luhman99}
Luhman, K.~L.\ 1999, \apj, 525, 466
\bibitem[Miglio \& Montalb{\'a}n(2005)]{Miglio05}
Miglio A. \& Montalb{\'a}n J. 2005, A\&A, 441, 615
\bibitem[Palla et al.(2005)]{Palla05}
Palla, F., Randich, S., Flaccomio, E., \& Pallavicini, R. 2005, ApJ, 626, L49
\bibitem[Palla et al.(2007)]{Palla07}
Palla et al. 2007, ApJ, submitted
\bibitem[Roxburgh \& Vorontsov(2003)]{Roxburgh03}
Roxburgh I.~W., Vorontsov S.~V., 2003, A\&A, 411, 215
\bibitem[Sacco et al.(2007)]{Sacco07}
Sacco et al. 2007, A\&A, in press (astro-ph/0611880)
\bibitem[Siess et al.(2000)]{Siess00}
Siess, L., Dufour, E., \& Forestini, M.\ 2000, \aap, 358, 593
\bibitem[Skumanich(1972)]{Skumanich72}
Skumanich, 1972, \apj, 171, 565
\bibitem[Stauffer(2004)]{Stauffer04} 
Stauffer, J.~R.\ 2004, ASPC, 324, 100 
\bibitem[Stauffer et al.(1995)]{Stauffer95} 
Stauffer, J.~R., Hartmann, L.~W., \& Barrado y Navascues, D.\ 1995, \apj, 454, 910 
\bibitem[Stauffer et al.(2003)]{Stauffer03} 
Stauffer, J.~R. et al.\ 2003, \aj, 126, 833 
\bibitem[Tassoul(1980)]{Tassoul80}
Tassoul, M.\ 1980, \apjs, 43, 469
\bibitem[Th{\'e}venin et al.(2002)]{Thevenin02}
Th{\'e}venin F., et al. 2002, A\&A, 392, L9
\bibitem[Thoul et al.(2003)]{Thoul03}
Thoul A., et al. 2003, A\&A, 402, 293
\bibitem[Timmes \& Arnett(1999)]{Timmes99} 
Timmes, F.~X., \& Arnett, D.\ 1999, \apjs, 125, 277 
\bibitem[White \& Hillenbrand(2005)]{White05} 
White, R.~J., \& Hillenbrand, L.~A.\ 2005, \apjl, 621, L65 
\bibitem[de Wit et al.(2006)]{deWit06}
de Wit, W. J., et al. 2006, A\&A, 448, 189
\end{thebibliography}
\end{document}